\documentclass[twocolumn,prl]{revtex4-1}
\usepackage{graphicx}
\usepackage{amsmath,amsthm,bbm,times}
\usepackage{amssymb}
\usepackage{bm}
\usepackage{bbold}
\usepackage{color}

\newtheorem{theorem}{Theorem}

\newtheorem{cor}[theorem]{Corollary}

 \def\idty{{\mathchoice {\mathrm{1\mskip-4mu l}} {\mathrm{1\mskip-4mu l}} %
{\mathrm{1\mskip-4.5mu l}} {\mathrm{1\mskip-5mu l}}}}
 \def\1{{\mathchoice {\mathrm{1\mskip-4mu l}} {\mathrm{1\mskip-4mu l}} %
		{\mathrm{1\mskip-4.5mu l}} {\mathrm{1\mskip-5mu l}}}}
		
\newcommand{\be}{\begin{equation}}
\newcommand{\ee}{\end{equation}}
\newcommand{\bea}{\begin{eqnarray}}
\newcommand{\eea}{\end{eqnarray}}
\newcommand{\beann}{\begin{eqnarray*}}
\newcommand{\eeann}{\end{eqnarray*}}
\newcommand{\eq}[1]{(\ref{#1})}

\newcommand{\cD}{{\mathcal D}}

\newcommand{\ket}[1]{\vert #1 \rangle}

\newcommand{\vp}{\varphi}

\newcommand{\gap}{\operatorname{gap}}
\newcommand{\supp}{\operatorname{supp}}

\newcommand{\spa}{\operatorname{span}}

\date{\today}

\begin{document}

\title{Low-complexity eigenstates of  a $\pmb{ \nu = 1/3}$ fractional quantum Hall system}
\author{Bruno Nachtergaele}
\affiliation{Department of Mathematics
and Center for Quantum Mathematics and Physics,
University of California, Davis, CA  95616-8633, USA}
\author{Simone Warzel}
\affiliation{Munich Center for Quantum Science and Technology, and
Zentrum Mathematik, TU M\"{u}nchen,
85747 Garching, Germany}
\author{Amanda Young}
\affiliation{Munich Center for Quantum Science and Technology, and
Zentrum Mathematik, TU M\"{u}nchen,
85747 Garching, Germany}

\begin{abstract}
We identify the  the ground-state of a truncated version of Haldane's pseudo-potential Hamiltonian in a thin cylinder geometry as being composed of exponentially many fragmented matrix product states. These states are constructed by lattice tilings and their properties are discussed. We also report on a proof of a spectral gap, which implies the incompressibility of the underlying fractional quantum Hall liquid at maximal filling $ \nu = 1/3 $. Low-energy excitations
and an extensive number of many-body scars at positive energy density, but nevertheless low complexity, are also identified using the concept of tilings.
\end{abstract}

\maketitle

\section{Introduction}

Our ability to make systematic approximations of physically relevant states of quantum many-body systems rests on the relatively low complexity of these states. A useful measure of complexity is given by bipartite entanglement. Although generic states of quantum many-body systems are nearly maximally entangled \cite{hayden:2006}, ground states of Hamiltonians with short-range interactions are expected to generically satisfy an area law for the bipartite entanglement, or an enhanced area law, and this has been proved for a number of important classes of systems \cite{cramer:2006,wolf:2006,hastings:2007}. The exact ground states of frustration free models provided by Matrix Product States (MPS) and Tensor Network States (TNS) \cite{fannes:1992,perez-garcia:2007} make
this low-complexity aspect explicit. The first examples of MPS and TNS were given in \cite{affleck:1987a,affleck:1988}, where they were called Valence Bond Solid states, a name which itself suggested low complexity and limited entanglement. The success of DMRG inspired numerical approaches is also best understood based on its connection with MPS and TNS \cite{cirac:2009,schollwock:2011}.

In this letter, we report on a recent detailed investigation of the ground state, spectral gap, and incompressibility of a popular model for the Fractional Quantum Hall Effect (FQHE) in a $1/3$ filled system: a truncated version of Haldane's pseudo-potential Hamiltonian \cite{PhysRevLett.51.605}, which is defined in the one-dimensional 
lattice and describes the thin cylinder regime. On an interval $ \Lambda =[a,b]  $ the Hamiltonian  is \cite{PhysRevLett.54.237,Pokrovsky_1985,Trugman:1985lv,Bergholtz_2006,Nakamura:2012bu} 
\begin{equation}\label{eq:Ham}
H_\Lambda =  \sum_{x= a}^{b-2}  n_x n_{x+2} + \kappa \sum_{x= a}^{b-3} q_x^* q_x  , 
\end{equation}
where 
$n_x := c_x^* c_x$,
$q_x := c_{x+1}  c_{x+2} -  \lambda \ c_{x}  c_{x+3}$, 
and $c_x$ is the annihilation operator for a spinless fermion at site $x$. Equivalently, the model can be regarded as a 
spin-$1/2$ chain. The parameters $\kappa$ and $\lambda$ are given in terms of the dimensionless parameter $\alpha=\ell/R$, the ratio
of the magnetic length and the radius of the cylinder geometry of the system: $\kappa = e^{3\alpha^2/2} / 4 $ and $ \lambda = -3 e^{-2\alpha^2} $.  
In what follows, we take model parameters $ \kappa > 0 $ and $ \lambda \in \mathbb{C}$ which are arbitrary unless otherwise stated. 

The Hamiltonian~\eqref{eq:Ham} is an example of a particle-number $ N = \sum_x n_x $ and dipole-conserving  $ \sum_x x n_x $  model. 
Its spectrum is invariant under $ \lambda \to - \lambda $ as is seen by unitarily transforming $ c_x  \to - c_x $ on every 4th site $ x $ of the chain.

In addition to proving a spectral gap, we investigate the excitation spectrum. Using the concept of tilings, which we use to classify an orthogonal basis for the ground state space, we identify invariant subspaces of the Hamiltonian and prove asymptotically sharp bounds on low-lying excitations. Many-body scars of higher energy but bounded Schmidt rank are also constructed using tiling states.

\section{Ground states -- tilings and fragmented MPS}

In the Tao-Thouless (TT) limit $ \lambda = 0 $ \cite{Tao:1983cd}, the ground-state at maximal filling $ \nu = 1/3 $ is given by the tensor state $ \ket{(100)_L} $ of $ 3$-periodic particle configurations which repeat $ L $-times the basic particle pattern (100) indicating the presence of a particle at every third site. A ground-state of \eqref{eq:Ham} with $ |\Lambda | = 3 \, L  $ at maximal filling is obtained by squeezing
\begin{equation}\label{eq:sTTstate}
\vp_L :=   \prod_{k = 0  }^{L-2} \left( 1 + \lambda c_{3k+2}^* c_{3k+3}^* \, c_{3k+4}  c_{3k+1}  \right)  \ket{(100)_L}  .
\end{equation}
Shifting the reference state $ (100)_L $ by 1 or 2 lattice sites produces two other ground states. 
The full ground-states space $ \ker H_\Lambda=  \{ \psi | H_\Lambda\psi = 0 \} $ contains exponentially many other states  of lower filling. 
The set of particle configurations, which support ground states, is constructed
by concatenating voids written as (0), and monomers (100). In addition, at the left and right boundary of the finite chain some additional configurations are allowed: (11000) at the left boundary and (011), (10), and (1) at the right boundary. Concatenating these basic patterns or {\em tiles} produces a {\em root tiling} and configurations obtained in this way are called {\em root configurations}. Each root tiling $ R $ corresponds a basis vector of $ \ker  H_\Lambda $. 
These vectors are superpositions of many-particle configurations, namely all those that correspond to a compatible tiling of the finite chain in which pairs of successive monomers are replaced with dimers, i.e.
\[  (100)(100) \to (011000) . \]
At the right boundary special replacement rules apply: $(100)(10) \to (01100)$ and $(100)(1) \to (0110)$. For every root tiling $R$ we denote by $ \mathcal{D}_\Lambda(R) $ the
compatible tilings with Voids, Monomers, and Dimers (VMD tilings, for short), obtained by making the allowed subsitutions as described above. The ground state vector corresponding to $R$ is defined as:
\begin{equation}\label{eq:VMDstate}
	\psi_\Lambda(R) := \sum_{\pmb{D}  \in  \mathcal{D}_\Lambda(R) } \lambda^{\#( \pmb{D} )}  \ket{  \pmb{\sigma}(\pmb{D}) } .
\end{equation}
The sum extends over the collection of VMD tilings  $ \pmb{D}  \in\cD_\Lambda(R)$. Each VMD tiling $  \pmb{D} $ is associated with unique particle configuration $  \pmb{\sigma}(\pmb{D})  $ and $ \#( \pmb{D} ) $ counts the number of dimers in the tiling.  

\begin{theorem}[Ground State Space]
For all $|\Lambda|\geq 8$, the set of $\psi_\Lambda(R)$ labeled by the root tilings $R$ is an orthogonal basis of the ground state space of $H_\Lambda$. The dimension of the ground state space grows as $a^{|\Lambda|}$, with $a\approx 1.47$.
\end{theorem}

For more detailed information on this and other properties of ground state spaces, its dimension, and factorization formulas presented next, see
\cite{nachtergaele:2020}. 

A point of interest is the maximal filling of $1/3$. In fact, it follows easily from tiling structure that the largest particle number for a ground state, $N_\Lambda^{\rm max}$, satisfies
\begin{equation}\label{eq:maxNumber}
	 \frac{1}{3} \leq \frac{ N_\Lambda^{\rm max}}{|\Lambda|} \leq  \frac{1}{3} + \frac{4}{3|\Lambda|} .
\end{equation}
Therefore, the maximal filling in the ground state is asymptotically equal to 1/3.

The squeezed TT-state~\eqref{eq:sTTstate} is the state from~\eqref{eq:VMDstate} generated by the root $ R= (100)_L $. This state has Schmidt rank 2 across any bipartition of the lattice, which is most easily seen through the recursion relation
\begin{equation}\label{eq:recrel} 
\varphi_{L_1+L_2} = \varphi_{L_1} \otimes  \varphi_{L_2} + \lambda \varphi_{L_1-1} \otimes | 011000 \rangle \otimes \varphi_{L_2-1} .
\end{equation}
The correlation length of this state in the infinite volume limit was estimated in~\cite{Jansen:2009gv,Jansen:2012da} and identified in~\cite{Nakamura:2012bu} as 
\[ \xi(\lambda) := 3/\ln \frac{ \sqrt{4|\lambda|^2 +1}+1}{\sqrt{4|\lambda|^2 +1}-1}. \]

Up to  boundary tiles $ B_l $ and $ B_r $, which give rise to states $ \psi_{\partial^l \Lambda}(B_l) $ and  $\psi_{\partial^r \Lambda}(B_r) $, 
any state $\psi_\Lambda(R)$ 
is a tensor product of squeezed TT-states interspersed with
sites where the state is $\ket{0}$. The latter occur where the root tiling has voids $(0) $. 
The state can thus be regarded as a concatenation of intervals of length $3L_j$ on which the state is  $\vp_{L_j}$, and intervals in between where the
state is the vacuum, i.e. 
\[
\psi_\Lambda(R) = \psi_{\partial^l \Lambda}(B_l) \otimes \varphi_{L_1} \otimes | 0 \rangle \otimes  \dots   | 0 \rangle   \otimes  \varphi_{L_n} \otimes  \psi_{\partial^r \Lambda}(B_r) .
\]
 We call this structure a {\em fragmented matrix product state} (FMPS) since each fragment $ \vp_L $ (as well as the vacuum and boundary states) is an MPS \cite{Nakamura:2012bu}.

One consequence of this factorization property, is the exponential decay of correlations $ \langle \cdot \rangle_{\Lambda,R} $ in each of the states $\psi_\Lambda(R)$ proven in~\cite{nachtergaele:2020}. 
\begin{theorem}[Exponential clustering] \label{thm:expcluster}
There are positive constants $C$ and $\xi\leq 2 \xi(\lambda) $ such that for all finite chains $\Lambda$ and root tilings $R$ the correlations in the state $\psi_\Lambda(R)$ satisfy
$$
|\langle AB\rangle_{\Lambda,R} - \langle A\rangle_{\Lambda,R} \langle B\rangle_{\Lambda,R}|\leq C \Vert A\Vert \Vert B\Vert e^{-\frac{d(\supp A, \supp B)}{\xi}} . 
$$
\end{theorem}

Another immediate consequence of the FMPS structure and~\eqref{eq:recrel}  is the uniform bound $ 2 $ on the Schmidt rank of $\psi_\Lambda(R)$  across any cut in the chain. (Hence the entanglement entropy is bounded by $\log 2$.)

In view of the results presented next, 
one could say that each of these states
has the properties expected of gapped ground states based on general results about exponential clustering~\cite{hastings:2006,nachtergaele:2006a} and areas law in one dimension \cite{hastings:2007,brandao:2015a}. 
At the same time, however, the ground state space also contains vectors with extensive entanglement and states with arbitrarily slow decay of correlations, 
cf. Sec. 4.3 in \cite{nachtergaele:2020}. Given the large degeneracy of the ground states, this is not surprising, but it does raise an interesting 
question about the excited states: is there a basis of low-complexity eigenvectors of the model at all 
energies or a particular range of energies? By constructing eigenstates that explain some prominent features of the low-lying energy
spectrum we obtain evidence in support of the existence of a low-complexity basis of excited states.

\section{Spectral gap and incompressibility}

\begin{figure}
\includegraphics[width=.4\textwidth]{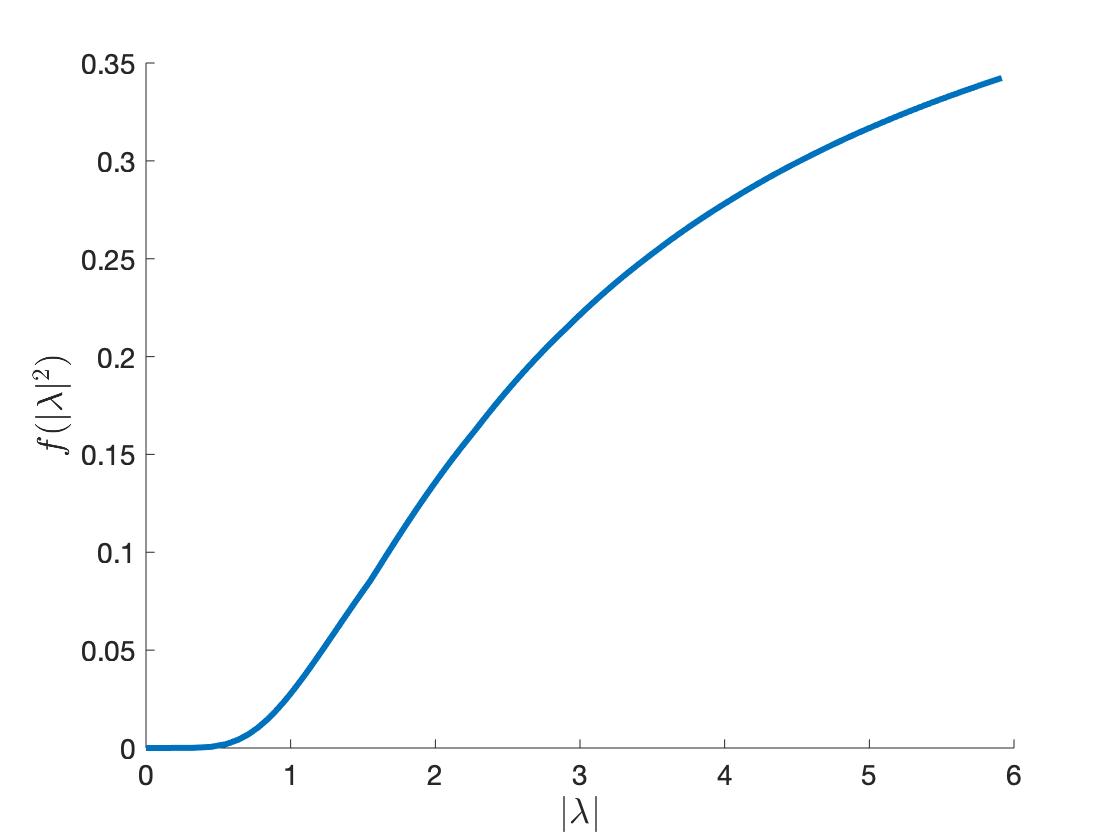}
\caption{Plot of the function $|\lambda|\mapsto f(|\lambda|^2)$.} \label{fig:f}
\end{figure}

An essential ingredient in the description of the FQHE is the gap in the spectrum above the ground state. In \cite{nachtergaele:2020} we proved the following lower bound for this gap which is uniform in $\Lambda $. The estimate is expressed in terms of the gap for chains of length $L=8,9,10$, which can be easily calculated numerically, and a function $f(|\lambda|^2)$. This function is defined and analyzed in  \cite{nachtergaele:2020}. Its essential properties can be read off from the plot given in Fig.~\ref{fig:f}.

\begin{figure}
\includegraphics[width=.5\textwidth]{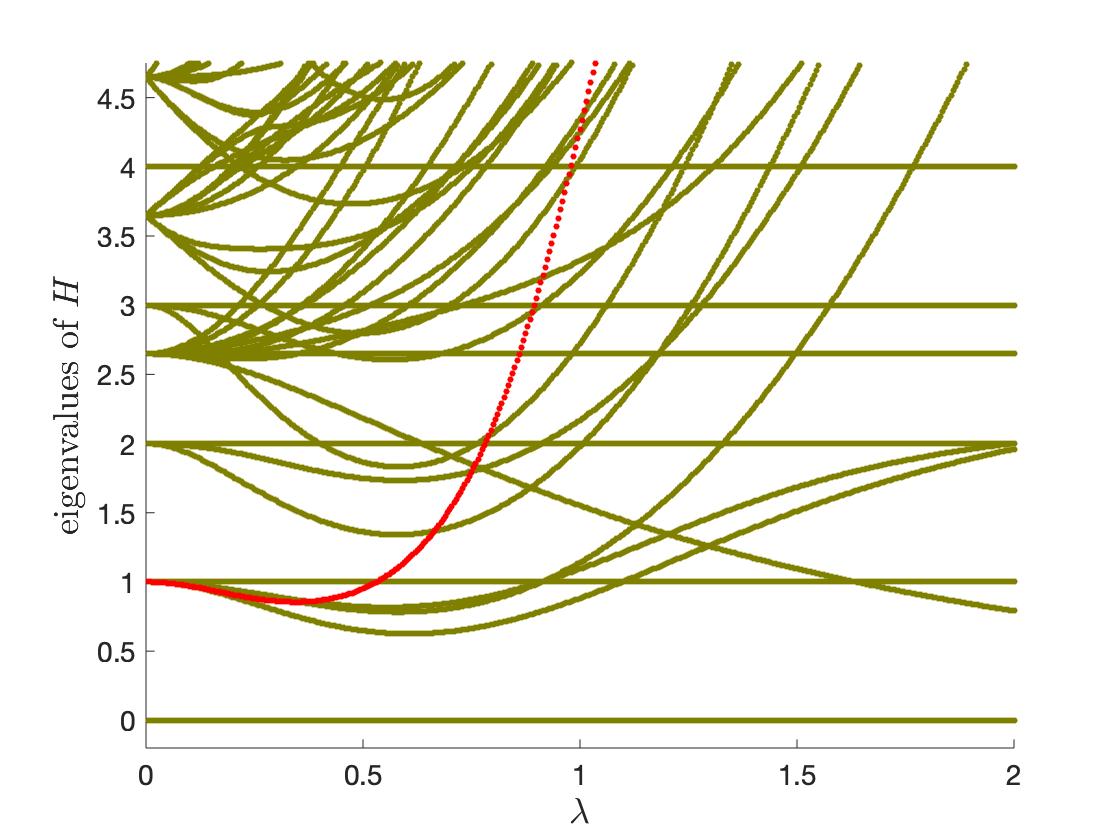}
\includegraphics[width=.5\textwidth]{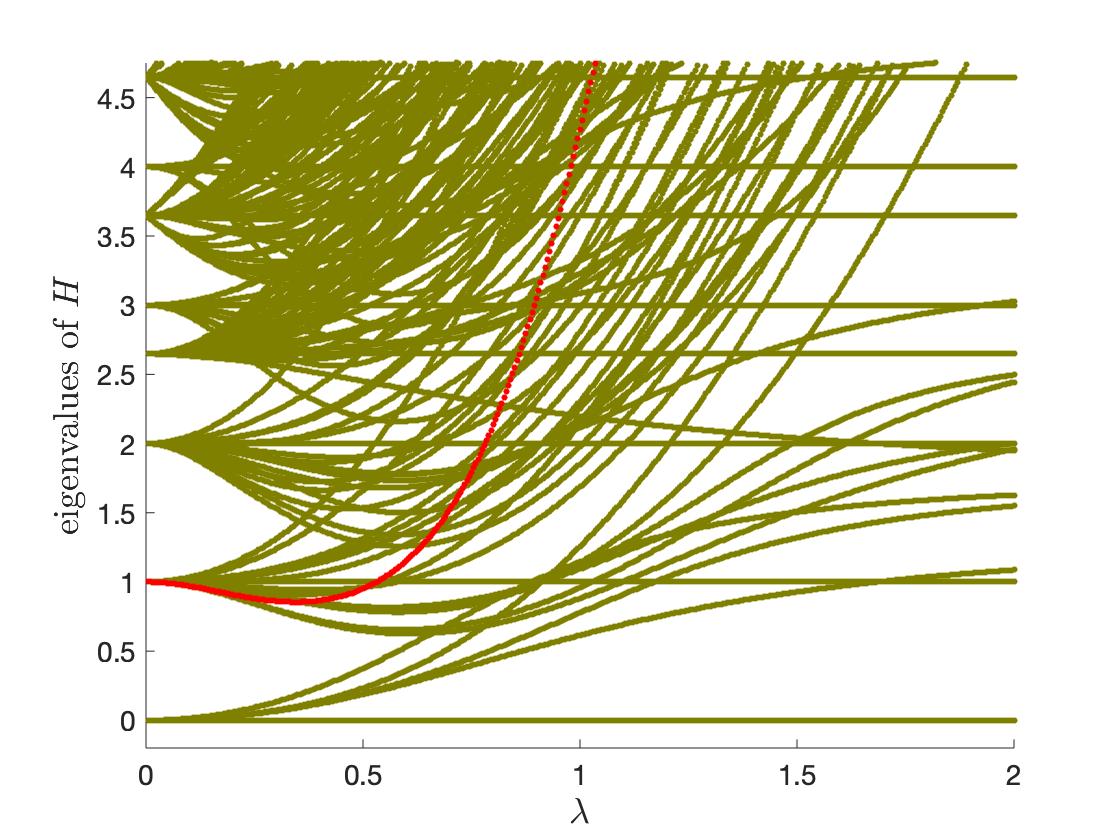}
\caption{Excitation spectrum as a function of $\lambda \in [0,2]$, for a chain of length $12$ with periodic (top) and open (bottom) boundary conditions with a fixed value of $\kappa = 2.648$. The red curve is the variational upper bound of the first excited state given in (\ref{eq:upper_bound}). } \label{fig:spec}
\end{figure}

\begin{theorem}[Uniform spectral gap]\label{thm:gap}
	There is a monotone increasing function $ f : [0,\infty) \to  [0,\infty) $ such that for all $ \lambda \in \mathbb{C} $ with the property $  f\left(|\lambda|^2\right)  < 1/3  $ and all $ \kappa > 0 $ 
	\begin{equation}\label{eq:gaplower}
\gap H_\Lambda \geq \left( \min_{L\in \{8,9,10\}} \gap H_{[1,L]} \right)   \frac{\left(1-  \sqrt{3 f\left(|\lambda|^2\right)}\right)^2}{3} 
	\end{equation}
	for any interval $\Lambda$ of length $L\geq 9$.
\end{theorem}
For open boundary conditions, the gap of \eq{eq:Ham} vanishes when $\lambda\to 0$, cf.\ Fig.~\ref{fig:spec}. This is caused by spurious boundary states (which are also discussed in~\cite{PhysRevB.85.155116})
such as the eigenstate with lowest energy $E=\mathcal{O}(|\lambda|^2) $ in the 2-dimensional Krylov space $ \spa \{ H_{\Lambda}^n \ket{11001}\otimes \ket{(0)_{|\Lambda| -5}} \} $.
The bulk gap is not expected to vanish as $\lambda\to 0$. We discuss the presumed lowest energy excitations in the next section. Finite systems with periodic boundary conditions show the behavior of the bulk: the gap does not vanish in the limit  $\lambda\to 0$ (see  Fig.~\ref{fig:spec}).

Denoting by $E_L(N)$ the lowest energy state of $H_\Lambda$ with particle number $N$ on an interval of length $|\Lambda| =L$, the compressibility is defined by
\begin{equation}
\frac{1}{\kappa_L(N)}  :=   L  \,  \frac{E_{L+1}(N) + E_{L-1}(N)- 2 E_L(N)}{(2\pi \ell^2 )^2} \, . 
\end{equation}
As a consequence of the gap estimate, for fixed $\lambda$ the compressibility vanishes as $c/L$ in the maximally filled state.

\begin{cor}[Incompressibility]\label{cor:incompress}
	At zero temperature and critical filling, i.e.~$ N =  N_\Lambda^{\rm max} $, the compressibility $ \kappa_{\Lambda}(N) $ vanishes in the thermodynamic limit $  |\Lambda| \to \infty $.
\end{cor}

\section{Excited states}

The lowest neutral excitation in the $ \nu = 1/3 $  FQHE system is predicted to be a bound-state (`magnetoroton') of a quasi-particle of fractional charge $ -1/3 $ with quasi-hole of charge $ 1/3 $ \cite{PhysRevLett.54.581,PhysRevLett.108.256807}. 
As suggested by the numerical analysis in \cite{Nakamura:2012bu,wang:2015} as well as Fig.~\ref{fig:spec}, this remains true for the truncated Hamiltonian~\eqref{eq:Ham} in the bulk.
In the TT-limit $\lambda =0$, the first excited energy state is  obtained by introducing a kink $ (010) $ into the list of monomers. To show that a state with such features persists for small $|\lambda|>0$ and $ \kappa > 1 $, we look at the Krylov spaces $\mathcal{K}_{L,R}^{(l)}$ generated by $\vp_L \otimes \ket{(010)_l} \otimes\vp_R$, where $L$ and $R$ count the number of monomers to the left and right of the kink region, in which $ (010) $ is repeated $ l \in \mathbb{N} $ times.  
Due to the domain wall created by $ (0) $ on the left, this space factorizes as $ \mathcal{K}_{L,R}^{(l)} = \vp_L \otimes \ket{0} \otimes \mathcal{K}_R^{(l)}  $ where 
\begin{equation}\label{eq:krylov}
\mathcal{K}_R^{(l)} = \spa \{ H_\Lambda^n \left(  \ket{ (100)_{l-1}(10) }\otimes \vp_R \right) \, \big| \, n\geq 0\} 
\end{equation} 
and we identify $\Lambda=[1,3(R+l)-1]$. The choice of $ l\in \mathbb{N}  $ corresponds to shifting the dipole moment. It was noted in~\cite{Nakamura:2012bu} (and~\cite{PhysRevLett.108.256807} for the full model), that the lowest excitation occurs at $ l = 2 $. 

It is natural to embed  $\mathcal{K}_R^{(l)}  $ in the space $\mathcal{T}_R^{(l)} $ spanned by the tiling configurations generated from applying the replacement rules to the root tiling $(100)_{l-1}(10)(100)_R$. When $ l = 1 $, the presence of $(10) $ generates one additional tile of length 8 obtained from the rule:
\begin{equation}\label{eq:tilings1}
(10)(100)(100) \to (10)(011000) \to (01101000).
\end{equation} 
The deformation is thus localized to the two monomers following (10) and any tiling to the right of this region is a regular monomer-dimer tiling. 
For $ l \geq 2 $ we get two additional tiles from the replacement rules:
\begin{equation}\label{eq:tilings2}
(100)(10)(100) \to  (01100)(100) \to (01011000).
\end{equation} 

To study the lowest excitation, we use that $\mathcal{T}_R^{(2)}$ is orthogonal to the ground states, so that minimizing the energy $E(\sigma,\tau)$ over $\sigma, \tau \in\mathbb{C}$ of the variational state
\begin{equation}\label{eq:approx_estate}
\Psi_{\sigma,\tau} =   \ket{10010}\otimes \left( \vp_R + \sigma \eta_R \right) + \tau \ket{01100}\otimes  \vp_R  \in \mathcal{T}_R^{(2)},
\end{equation} 
produces an upper bound on the lowest excited energy. 
Here $\eta_R := -\bar{\lambda} \alpha_{R-1}^{1/2} \ket{100}\otimes \vp_{R-1} +  \alpha_{R-1}^{-1/2}\ket{011000}\otimes \vp_{R-2}$ has norm $ \| \eta_R\| = \| \vp_R \| $ and is orthogonal to $\vp_R= \ket{100}\otimes \vp_{R-1}  + \lambda   \ket{011000}\otimes \vp_{R-2}$ and 
\begin{equation}\label{eq:alpha}
\alpha_R := \frac{ \|\vp_{R-1}\|^2}{\|\vp_R\|^2} \stackrel{R\to \infty}{ \longrightarrow}  \frac{2}{1+\sqrt{1+4|\lambda|^2}} .
\end{equation}
Using~\eqref{eq:recrel}, one finds that the Schmidt-rank of  $\psi_{\sigma,\tau}$ is at most $4$ for any bipartition.
For $ \kappa > 1 $ and small $ |\lambda| $, where $ a :=  \alpha_R(1+\kappa\alpha_{R-1}|\lambda|^4) + \kappa |\lambda|^2 < \kappa $,  the variational minimum of the energy is asymptotically attained at
\begin{align*}
\sigma_{\min} & := \frac{\lambda  \alpha_R   \alpha_{R-1}^{1/2} (1-\kappa|\lambda|^2)}{b-a} =  \frac{\lambda}{\kappa-1} \left(1+\mathcal{O}(|\lambda|^2)\right) \\
\tau_{\min}  & := \frac{\kappa \lambda}{c - a}   =  \frac{\kappa \lambda}{\kappa-1} \left(1+\mathcal{O}(|\lambda|^2)\right)
\end{align*} 
where $ b := \kappa + \kappa(1+\alpha_R+\alpha_{R-1}) |\lambda|^2 + \alpha_R\alpha_{R-1}|\lambda|^2 $  
and $ c := \kappa + \kappa \alpha_R |\lambda|^2 $. This yields:
\begin{align}\label{eq:upper_bound}
\min_{\Psi\in \mathcal{T}_R^{(2)}} & \frac{\langle \Psi , H_{\Lambda} \Psi \rangle}{\| \Psi \|^2}  \leq  \  E(\sigma_{\min} ,\tau_{\min}) \nonumber \\
= & \ a -  \frac{(b-a)  |\sigma_{\min}|^2 + (c-a)   |\tau_{\min}|^2 }{ 1+  | \sigma_{\min}|^2 +   |\tau_{\min}|^2} . 
\end{align}
This bound shares the asymptotic parabolic behavior  $ 1 - \frac{2\kappa}{\kappa-1} |\lambda|^2 + \mathcal{O}(|\lambda|^4) $
exhibited by the first exited state in Fig.~\ref{fig:spec} for small $|\lambda| $.

To show the bound in~\eqref{eq:upper_bound} is sharp up to $ \mathcal{O}(|\lambda|^2) $, we note that $\mathcal{T}_R^{(2)}$ is invariant under both $ H_{\Lambda}$ and  its lower bound $H_{[1,11]}$. 
The latter is block diagonal in the configuration basis of $\mathcal{T}_R^{(2)}$ with all non-trivial blocks given by one of two matrices, $h_6$ and $h_3$. The lowest eigenvalue belongs to the $6 \times 6$-matrix $h_6$ which is obtained from the action of $H_{[1,11]}$ on the 6 tiling configurations generated from $(100)(10)(100)_2$, see \eqref{eq:tilings1}-\eqref{eq:tilings2}. 
The minimal energy $E$ can then be seen to coincide with~\eqref{eq:upper_bound} up to second order $ \mathcal{O}(|\lambda|^2) $, e.g., by using the characteristic polynomial $p(x) = \det[h_6-(x+1+2\kappa|\lambda|^2)\idty]$ to bound
\begin{equation}
\min_{\Psi\in \mathcal{T}_R^{(2)}} \frac{\langle \Psi , H_{\Lambda} \Psi \rangle}{\| \Psi \|^2}  \geq E \geq 1+2\kappa|\lambda|^2-\frac{p(0)}{p'(0)}.
\end{equation}

The same procedure can be applied to asymptotically estimate the minimum of the spectrum of $ H_\Lambda $ on other sectors of quasi-momenta. E.g., on $ \mathcal{T}_R^{(1)}$ one can similarly show that $ \min_{\Psi\in \mathcal{T}_R^{(1)}} \frac{\langle \Psi , H_{\Lambda} \Psi \rangle}{\| \Psi \|^2}  = 1 - \frac{\kappa}{\kappa-1} |\lambda|^2 + \mathcal{O}(|\lambda|^4) $, which is clearly greater than~\eqref{eq:upper_bound}.\\

Bulk excitations of low complexity other than the magnetoroton can be constructed by decoupling a finite-volume excited state that is separated by pair of domain walls from two ground states. 
E.g., the eigenvalues in Fig.~\ref{fig:spec} which are constant in $ \lambda $ are constructed by putting a void $ (0) $ on both sides of $(10)(100)$. An excited state with energy $E=1$ is then given by
\begin{equation}
\psi = \vp_{L} \otimes \ket{0101000} \otimes \vp_{R} . 
\end{equation}
Similarly, placing voids around each of $N$-pairs $(10)(100)$ and extending the state to the full system by tensoring with appropriate ground states produces an excited state with energy $N$. Such states all have Schmidt-rank~$ 2 $.

Other excited states can similarly be generated as long as their boundary does not resonate with the domain wall. However, the minimal number of voids needed to construct a sufficiently large domain wall can vary. For example, the highest energy state $\ket{1}^{\otimes N}$ on $N$ sites requires one void on the left and three voids on the right:
\[
\psi = \vp_L \otimes \ket{0111\ldots 1000} \otimes \vp_R.
\]
This produces an eigenstate with energy $ N-2+(N-1)\kappa + (N-3)\kappa|\lambda|^2$.

\section{Discussion}
Motivated by the above description of the ground-state and the magnetoroton, we conjecture the existence of a complete set of low-energy eigenstates with low-complexity.
This would be in accordance with the hierarchical construction of excited states in this model~\cite{Bergholtz_2006}; see also \cite{PhysRevLett.51.605,PhysRevLett.52.1583}.

Our construction of strictly local eigenstates at medium of high-range in energies is akin to that of the `frozen states' 
in related dipole-conserving models~\cite{Pollmann19b}. For the model in which one drops in~\eqref{eq:Ham} all electrostatic interactions, a comprehensive discussion of such eigenstates is found in~\cite{arXiv:1906.05292v1,Bernevig10.19}.  Such eigenstates are known as many-body scars and have been identified recently in many other systems \cite{PhysRevB.101.174308,PhysRevB.101.195131}. They violate the Eigenfunction Thermalization Hypothesis (ETH):
although they share common global quantum numbers, 
expectations of local observables differ. Nevertheless, the model is not integrable nor is it, as far as we know described by Local Integrals of Motion (LIOMs), as in case of many-body localized systems.

The states discussed here are always of low-complexity and hence should be easily experimentally accessible. The description is the guided by the same principle employed for the ground states and in~\cite{Bernevig10.19}. 
Using a root-tiling we generate a set of configurations (and hence tensor-product states) via replacement rules that constitutes an invariant subspace of the Hamiltonian.

\begin{acknowledgments}
This work was supported by the DFG (Germany) under EXC-2111--390814868 and by the National Science Foundation (USA) under Grant DMS-1813149.
\end{acknowledgments}

%
\end{document}